\documentclass[a4paper]{PoS}

\usepackage{epsf,amssymb,amscd}
\usepackage{shuffle}

\newcommand{\be}{\begin{equation}}
\newcommand{\ee}{\end{equation}}
\newcommand{\bea}{\begin{eqnarray}}
\newcommand{\eea}{\end{eqnarray}}
\newcommand{\beas}{\begin{eqnarray*}}
\newcommand{\eeas}{\end{eqnarray*}}

\def\One{\mathbb{I}}

\def\wthree{\;\raisebox{-3mm}{\includegraphics[height=8mm]{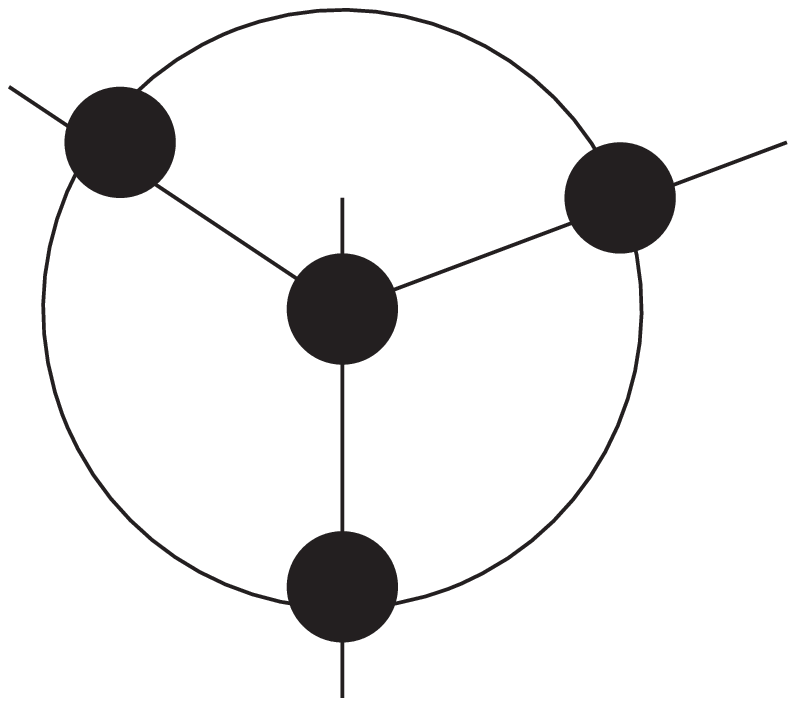}}\;}
\def\wfour{\;\raisebox{-3mm}{\includegraphics[height=8mm]{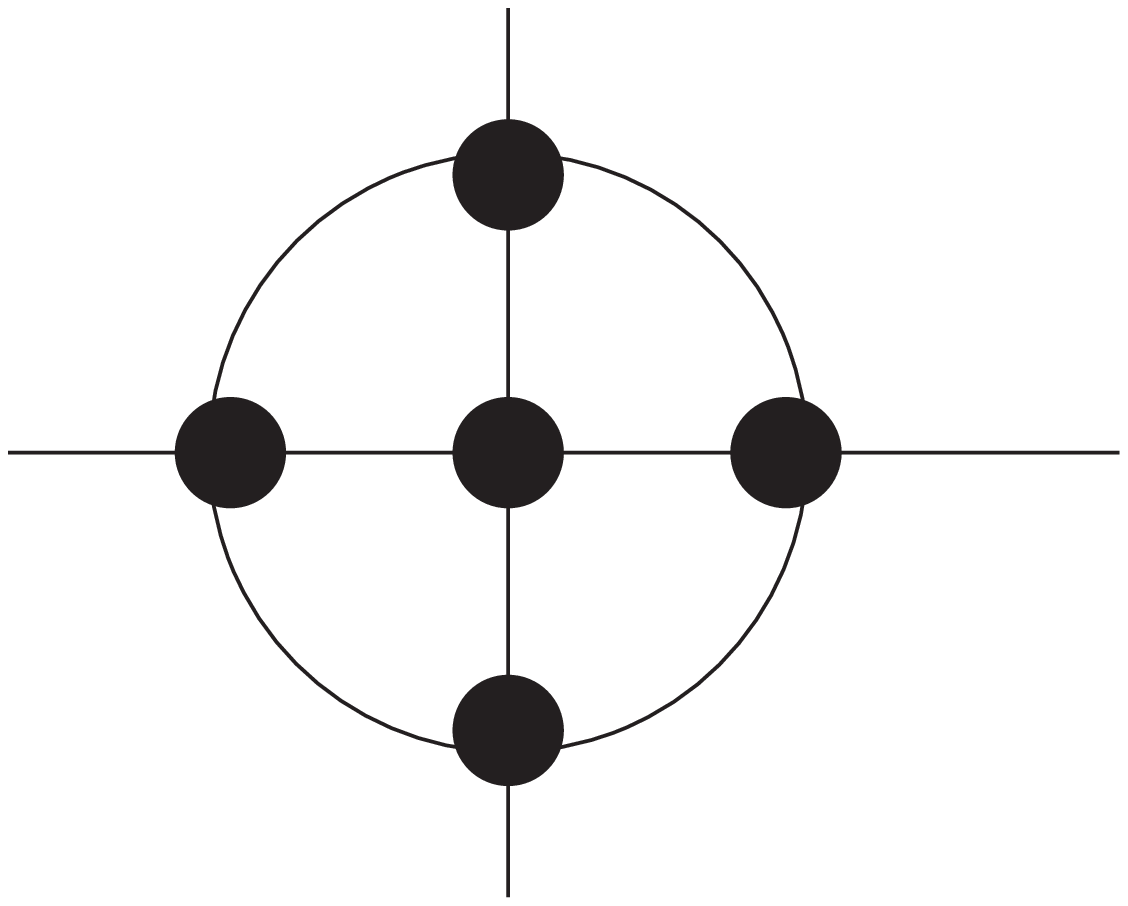}}\;}
\def\wft{\;\raisebox{-4mm}{\includegraphics[height=12mm]{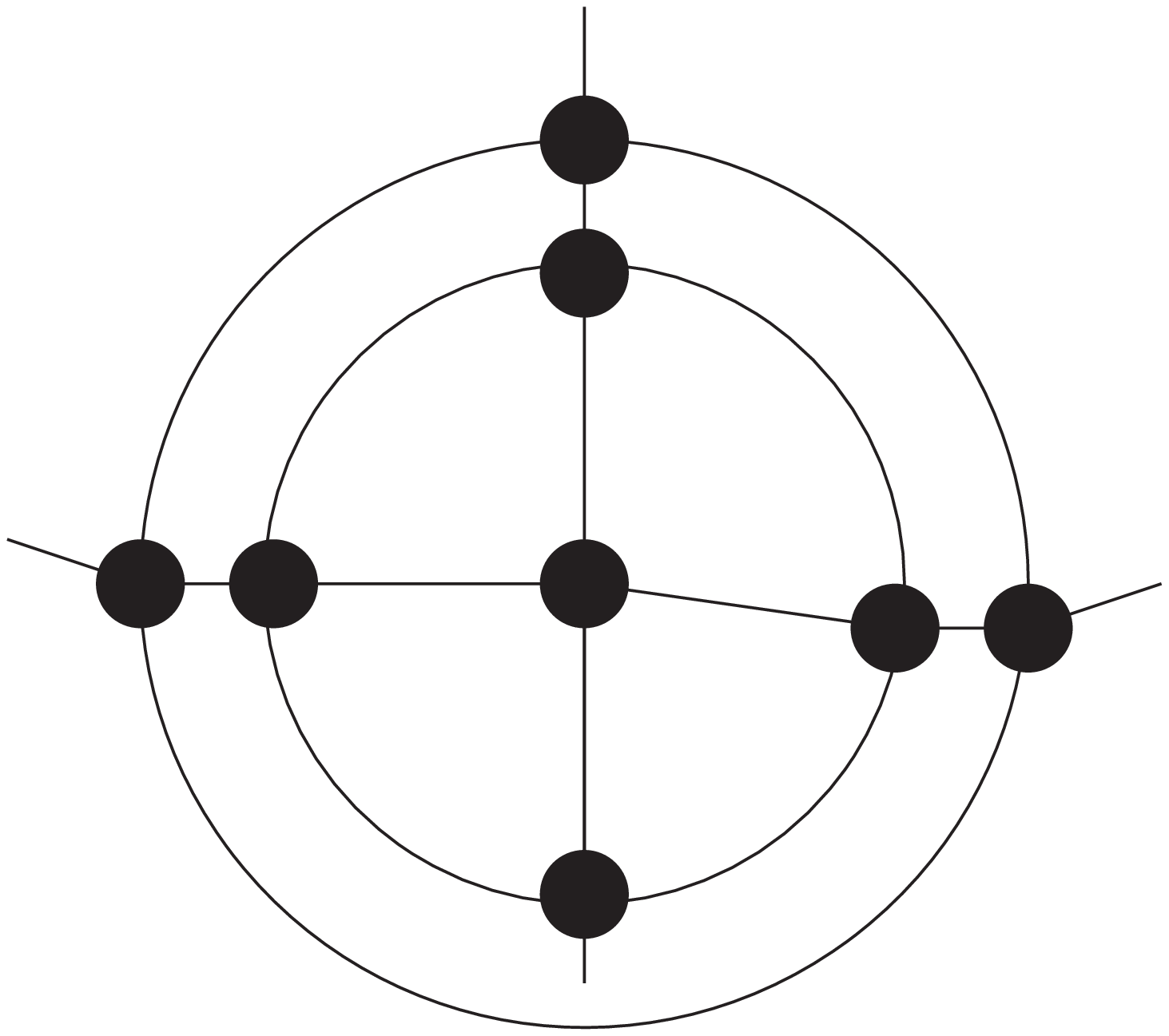}}\;}
\def\wtfa{\;\raisebox{-4mm}{\includegraphics[height=12mm]{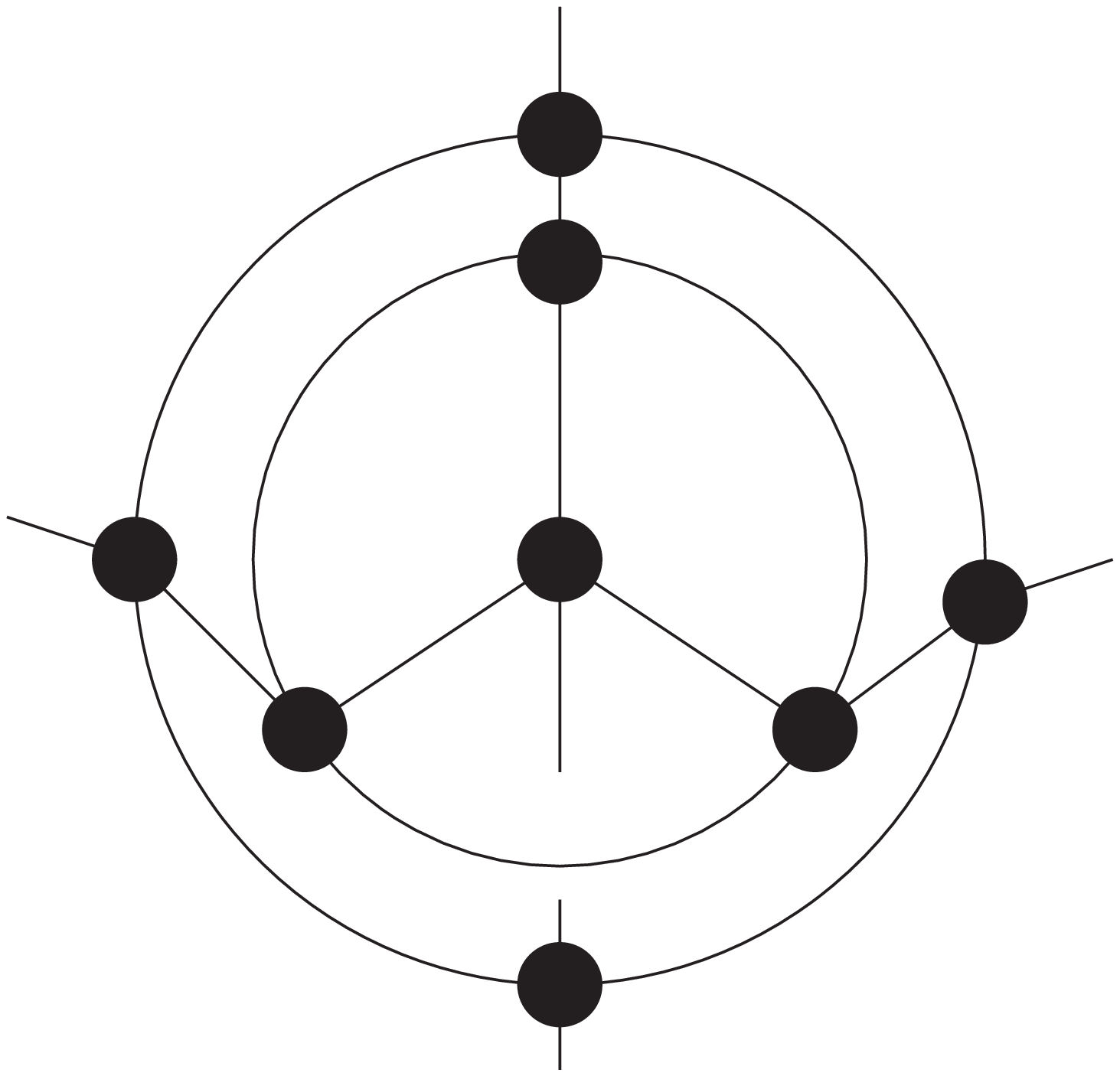}}\;}
\def\wtfb{\;\raisebox{-4mm}{\includegraphics[height=12mm]{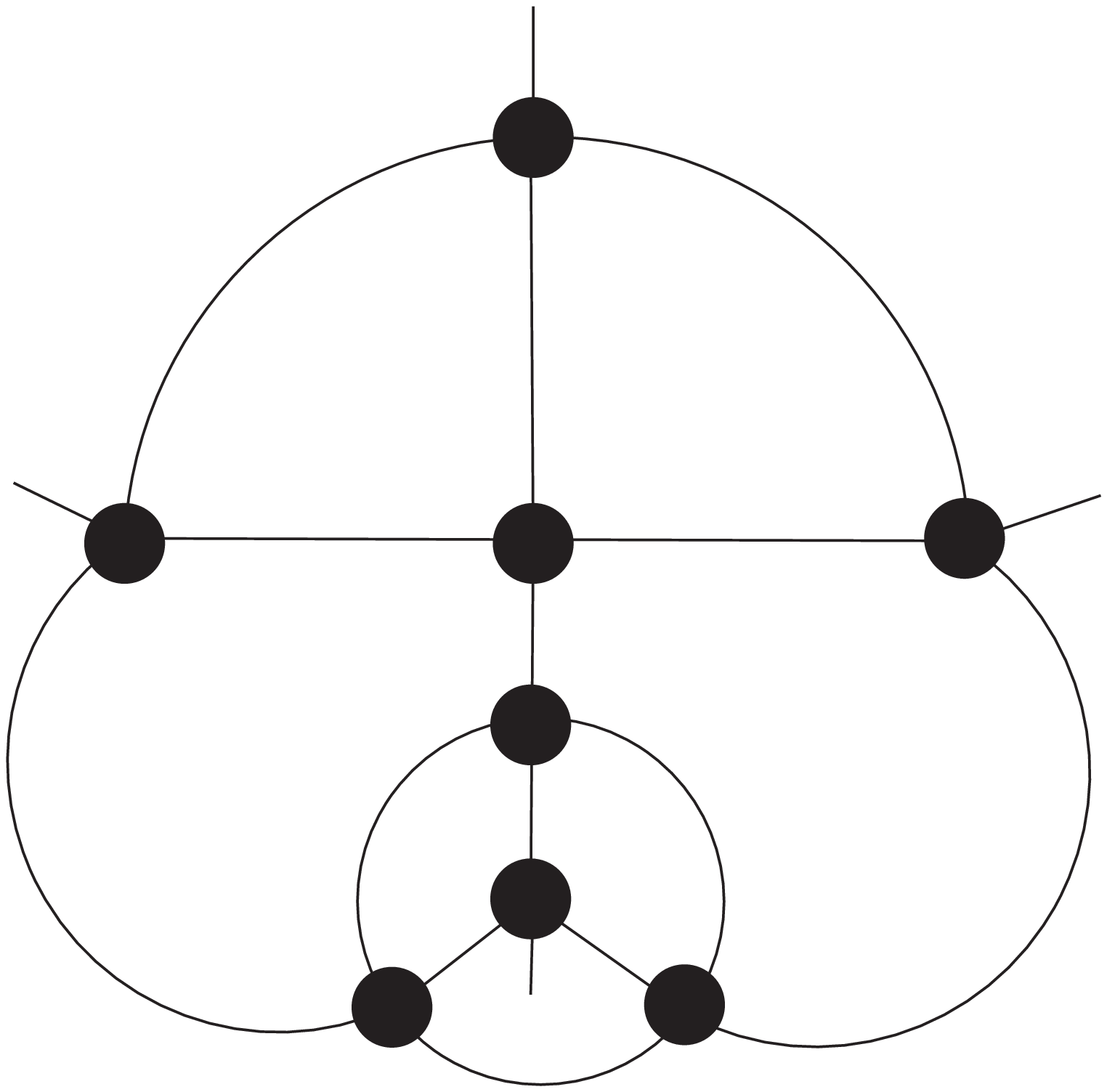}}\;}

\title{What can we learn from Knizhnik--Zamolodchikov Equations?}

\ShortTitle{What can we learn from Knizhnik--Zamolodchikov Equations?}

\author{\speaker{Dirk Kreimer}%
         \thanks{Author supported by the Alexander von Humboldt Foundation and the BMBF through an Alexander von Humboldt Professorship.}\\
        Humboldt Universit\"at Berlin\\
        E-mail: \email{kreimer@physik.hu-berlin.de}}

\abstract{
We discuss structural similarities between Knizhnik--Zamolodchikov equations (in fact, their simplest version needed to introduce the Drinfel'd associator)
and Dyson--Schwinger equations. We emphasize that the latter allow for a filtration by co-radical degree using quasi-shuffle products and the lower central series filtration of the Lie algebra of Feynman graphs. This clarifies how they are a generalization of the KZ equations. This is a starting point for a algebraic organization of the next-to$\cdots$-to leading log expansion which has been worked
out in collaboration with Olaf Kr\"uger and which will be given elsewhere \cite{OKMaster,KrKr}.
}
\FullConference{ Loops and Legs in Quantum Field Theory - LL 2014,\\
                 27 April - 2 May 2014 \\
                 Weimar, Germany }

\begin{document}
\section{The Knizhnik Zamolodchikov Equation}
Let us quickly remind ourselves of the most basic set-up for the KZ equations.
\begin{itemize}
\item We consider a free Lie  algebra $L$ of words $w$, in a two-letter alphabet on letters $a,b$, and assign a co-product 
$$\Delta w = \sum_{uv=w}u\otimes v\equiv \sum_{u\subset w}u\otimes w/u$$ to such words. We allow for $u$ or $v$ to be the full word $w$, and identify the $\emptyset$
with the unit $\One$, $\One w=w\One=w$.

\item Evaluation of iterated integrals: we can assign to such words iterated integrals in two differential forms, say $dz/z\leftrightarrow a$ and $dz/(1-z)\leftrightarrow b$, 
which we assign to the two letters $a,b$. This gives a natural map $\phi: L\to\mathbb{C}$ such that 
$$\phi(w_1\shuffle w_2)=\phi(w_1)\phi(w_2),$$ the evaluation of words is an algebra homomorphism for the shuffle algebra.
\item The KZ equation for us simply is 
$$ \frac{dF(z)}{dz}=\left(\frac{a}{z}+\frac{b}{1-z}\right)F(z).$$ The differential forms in it have poles at $0,1,\infty$. It pays to compare solutions regular at $0$ with solutions regular at $1$.
\item The associator $\Phi$ compares solutions regular at $0$ with solutions regular at $1$, and is hence a constant series in multi-commutators:
$$ \Phi=1+\zeta(2)[a,b]+\zeta(3)\left([a,[a,b]]-[b,[a,b]]\right)+\ldots.$$
\end{itemize}
Note that the KZ equation is a linear differential equations. It evaluates differential forms in a manner such that the shuffle product structure is preserved. 
\section{Generalized version}
We now want to generalize the above set-up so as to be flexible enough to incorporate Dyson--Schwinger equations (DSE).
For them, we still have a Hopf algebra structure, an associated Lie algebra structure such that the dual of the Hopf algebra of graphs is the universal enveloping algebra 
of that Lie algebra, and we have Feynman rules and the renormalization group equation (RGE). These structures combine so that we can interpret DSE as generalized KZ equations, by allowing for non-linearity and quasi shuffles. We need 
\begin{itemize}
\item A suitable  Lie  algebra $\mathcal{L}$ of graphs , with skeleton graphs -graphs free of subdivergences- playing the r\^ole of countably many letters $a,b,\cdots$.
\item A generalization to non-linear KZ equations, for example for the combinatorial DSE
$$X(g^2)=1-g^2B_+\left(\frac{1}{X(g^2)}\right),$$ we can set 
$$ \frac{dF(z)}{dz}=-g^2 a \left(\sum_{j=0}^\infty F^{\shuffle_\Theta j}\right)\frac{dz}{z}$$ as a generalized KZ equation (which was solved for Yukawa theory in \cite{BroadKr}).
The RGE ensures that for renormalized Feynman rules $\Phi_R$, $$\Phi_R(w_1\shuffle_\Theta w_2)=\Phi_R(w_1)\Phi_R(w_2),$$ 
which expresses the familiar fact that the leading logs are determined by the renormalization parts, while the non-leading terms can be captured via multi-commutators (see below)
or  in terms of a quasi-shuffle product
$$au_1\shuffle_\Theta bu_2=a(u_1\shuffle_ \Theta bu_2)+b(au_1\shuffle_\Theta u_2)+\Theta(a,b)(u_1\shuffle_\Theta u_2),$$
with $\Theta$ a commutative and associative map which assigns a new letter to any pair of letters.

The sum over Feynman graphs which appears as a solution of a combinatorial DSE then dualizes to a series in the dual universal enveloping algebra $\mathcal{U}(\mathcal{L})$ for a Lie algebra $\mathcal{L}$. 
Terms of highest order in the leading log expansion -elements of maximal coradical degree- correspond  to highest symmetric powers in $\mathcal{U}(\mathcal{L})$, while the linear terms (in $\ln S/S_0$)
are dual to elements of $\mathcal{L}\subset\mathcal{U}(\mathcal{L})$, and can be filtered themselves according to the lower central series filtration of $\mathcal{L}$, such that angle dependence is relegated to commutators, as in the example below  (see also \cite{Madrid} for a review of these properties of field theory). 

\item This formally gives generalized associators involving multi-commutators and images of  $\Theta$:
$$ \sim 1+q_1\frac{1}{2}\zeta(2)[a,b]+q_2\Phi\Theta(a,b)+\ldots, q_i\in\mathbb{Q}.$$ So what is $\Theta(a,b)$? A general  answer is given in \cite{KrKr}, based on the analysis in \cite{Madrid},
and an example is below.
\item For full DSE we have  iterated integrals of one forms replaced by renormalized Feynman rules for graphs with subgraphs. An exhaustive classification of Dyson--Schwinger equations as combinatorial fixed point equations and generators of sub-Hopf algebras, covering all known applications in physics, has been given by Lo\"ic Foissy \cite{Foissy}.
\item In analogy to KZ, the question then is what is the structure of $\sum_\Gamma\Phi_R(\Gamma)\Gamma$? 
\end{itemize}
\section{Hall series}
A crucial ingredient to tackle such questions is the study of Hall series.
\begin{itemize}
\item Start with the Lie algebra $\mathcal{L}$ of graphs $\Gamma$ whose universal enveloping algebra $\mathcal{U}(\mathcal{L})$ is the dual of a Hopf algebra $H$  with coproduct 
$$\Delta \Gamma = \sum_{\gamma\subset \Gamma}\gamma\otimes \Gamma/\gamma.$$
\item It has a lower central series filtration:
$\mathcal{L}$ acquires a descending series
of sub-algebras $ \mathcal{L} = \mathcal{L}_1 \unrhd \mathcal{L}_2
\unrhd \mathcal{L}_3 \unrhd \ldots $, where $\mathcal{L}_{n+1}$ is
generated by all $[x,y]$ with $x \in \mathcal{L}$ and $y \in
\mathcal{L}_n$.
\item For this, there is a Hall basis: lexicographical ordering of all elements in $\mathcal{L}$, for
example, let $x_1 < x_2 < \ldots < [x_1,x_2]< \ldots$ (it does not
matter which ordering we take as long as we choose one). We then
define $[x,x']$ to be a (Hall) basis element of $\mathcal{L}$ iff
both,
\begin{enumerate}
\item $x, x' \in \mathcal{L}$ are (Hall) basis elements with $x < x'$,
\item if $x' = [x'',x''']$, then $x\geq x''$
\end{enumerate}
are fulfilled.
\item This then provides next-to $\cdots$ to leading log expansions filtered by 'quasi'-ness ($\Theta$) and multi-commutators \cite{KrKr}.
\end{itemize}
We now mainly want to exhibit an example. For that, we have to first consider the structure of Green functions in QFT.
\section{Structure of a Green function}
Here, we summarize the results of \cite{BlKr,BrKr}.
\begin{itemize}
\item Our first concern is the decomposition of variables into couplings $g$ and kinematical variables $L,\theta,\theta_0$.
Here, $L=\ln S/S_0$ fixes a scale renormalized at $S_0$, dimensionless variables $\theta,\theta_0$ ('angles') are provided by scalar-products of external vectors or mass squares measured in units of $S$ or $S_0$ to completely specify the kinematics and renormalization conditions. 
\be G^R(\{g\},L,\{\theta,\theta_0\})=1\pm \Phi^R_{L,\{\theta,\theta_0\}}(X^r(\{g\}))\ee
with $$X^r=1\pm \sum_j g^j B_+^{r;j}(X^r Q^j(g)),$$ $bB_+^{r;j}=0$. This is 
\item Hochschild closedness:\\ '$B_+$ acts as it would append a first letter.'
\item
Then, for kinematic renormalization schemes, the group law $\star$ of the Hopf algebra of graphs is compatible with the additive character of scale variables $L$:
\item { $ \Phi^R_{L_1+L_2,\{\theta,\theta_0\}}=\Phi^R_{L_1,\{\theta,\theta_0\}} \star \Phi^R_{L_2,\{\theta,\theta_0\}}
$}. This is the RGE (\cite{BlKr,BrKr}).
\item We have an angle and scale separation (\cite{BrKr}):
$$\Phi^R(L,\{\ \theta,\theta_0\})=\Phi_{\mathrm{fin}}^{-1}(\{\theta_0\}\star\Phi^R_{\mathrm{1-scale}}(L)\star \Phi_{\mathrm{fin}}(\{\theta\}).$$ 
\end{itemize}

In general, in these variables, we have renormalized Feynman rules as forest sums
using Symanzik polynomials $\psi,\phi$:
\begin{eqnarray*}
 & & \Phi^R_\Gamma (L,\{\theta,\theta_0 \})
  =  \int_{\mathbb{P}^{E-1}(\mathbb{R}_+)}
\overbrace{\sum_f}^{\mathrm{forest sum}} (-1)^{|f|}
   \frac{\ln\frac{
\frac{S}{S_0}\phi_{\Gamma/f}\psi_f+\phi_f^0\psi_{\Gamma/f}}{\phi^0_{\Gamma/f}\psi_f+\phi_f^0\psi_{\Gamma/f}}}{\psi^2_{\Gamma/f}\psi^2_f} \underbrace{\Omega_\Gamma}_{(E-1)-\mathrm{form}}
\end{eqnarray*}
where we now have written scales and angles as arguments, not as subscripts.
These are well defined thanks to the marvellous properties of graph polynomials which spill over also to non-scalar theories \cite{BrKr,KrSavS}.\\
Note that $\Phi^R_\Gamma$ is a polynomial in $L$ when evaluated on any finite graph $\Gamma$,
and that the term linear in $L$ is given as
\begin{eqnarray*}
 & & \Phi^{R,1}_{\Gamma} (\{\theta,\theta_0 \})
  =  \int_{\mathbb{P}^{E-1}(\mathbb{R}_+)}
\sum_f (-1)^{|f|}
   \frac{1}{\psi^2_{\Gamma/f}\psi^2_f} \frac{\phi_{\Gamma/f}\psi_f}{\left(\phi_{\Gamma/f}\psi_f+\phi_f^0\psi_{\Gamma/f}\right)}\Omega_\Gamma
\end{eqnarray*}
We factored a scale $S/S_0$ from all second Symanzik polynomials.\\
But how do we finally get some quasi-shuffle $\Theta$? 

\section{Periods for co-commutative elements}
Here, we present an example which is also studied in \cite{Madrid}.
We consider the following (combination of) graphs in $\phi^4_4$.
\[\Gamma_3=\wthree,\;\Gamma_4\wfour,\;= \Gamma_{43}=\wft,\; \Gamma_{34a}=\wtfa,\; \Gamma_{34b}=\wtfb.\]
$$s_{34}=\Gamma_{43}+\frac{1}{2}\left( \Gamma_{34a}+\Gamma_{34b} \right)$$
$$c_{34}=\Gamma_{43}-\frac{1}{2}\left( \Gamma_{34a}+\Gamma_{34b} \right)$$
$$p_{34}=\Gamma_{43}+\frac{1}{2}\left( \Gamma_{34a}+\Gamma_{34b}-\Gamma_3\Gamma_4 \right)$$
$$\Delta(s_{34})=s_{34}\otimes\One+\One\otimes s_{34}+\wthree\otimes\wfour+\wfour\otimes\wthree $$
$$\Delta(c_{34})=c_{34}\otimes\One+\One\otimes c_{34}+\wthree\otimes\wfour-\wfour\otimes\wthree $$
$$\Delta(p_{34})=p_{34}\otimes\One+\One\otimes p_{34}.$$
\begin{itemize}
\item
Look at $s_{34}$ which is of co-radical degree two. Clearly, 
\[
\Phi_R(s_{34})=c_3^1c_4^1 L^2+c_{s_{34}}^1L.
\]
\item
In general, $c_{s_{34}}^1\equiv c_{p_{34}}^1$ is not a period but rather a complicated function of $\theta,\theta^0$.
\item
Assume that we subtract at $\theta=\theta_0$. $c_{s_{34}}^1$ could still depend on $\theta$. Alas
\begin{eqnarray*}
c_{s_{34}}^1 & = &  
 \int_{\mathbb{P}_\Gamma}\left( 
 \frac{1}{2} \frac{1}{\psi_{\Gamma_{34a}}^2}+\frac{1}{2} \frac{1}{\psi_{\Gamma_{34b}}^2}+\frac{1}{\psi_{\Gamma_{43}}^2}\right.\nonumber\\
  & & \left. 
-\frac{\phi_{\Gamma_3}\psi_{\Gamma_4}+\phi_{\Gamma_4}\psi_{\Gamma_3}}{\psi_{\Gamma_4}^2\psi_{\Gamma_3}^2\left[\phi_{\Gamma_3}\psi_{\Gamma_4}+\phi_{\Gamma_4}\psi_{\Gamma_3} \right]}\right)\Omega_\Gamma\nonumber\\ & = &
 \int_{\mathbb{P}_\Gamma}\left( 
 \frac{1}{2} \frac{1}{\psi_{\Gamma_{34a}}^2}+\frac{1}{2} \frac{1}{\psi_{\Gamma_{34b}}^2}+ \frac{1}{\psi_{\Gamma_{43}}^2}-\frac{1}{\psi_{\Gamma_4}^2\psi_{\Gamma_3}^2}\right) \Omega_\Gamma.
\end{eqnarray*}
All angle dependence has been eliminated.
\end{itemize}

\begin{itemize}
\item This is then $\Phi^R\circ\Theta(w_3,w_4)$, simply a new period which represents the new letter $\Theta(w_3,w_4)$. Erik Panzer promises me that his methods \cite{Erik} will produce this number in due time. 
\item All completely symmetric insertions of primitive graphs result in integrands involving only first graph polynomial and define periods corresponding to such new letters.
\end{itemize}
It remains to consider anti-symmetric insertions of graphs, corresponding to (multi-)commutators.
\section{Angle dependence in commutators}
\begin{itemize}
\item
For anti-cocommutative elements like $c_{34}$ angle dependence remains, \[\Phi_R(c_{34})=c_{34}^1(\theta)L.\]
\item  A simple computation reveals 
\begin{eqnarray}
c_{34}^1(\theta) & = &  
 \int_{\mathbb{P}_\Gamma}\left( 
 \frac{1}{2} \frac{1}{\psi_{\Gamma_{34a}}^2}+\frac{1}{2}\frac{1}{\psi_{\Gamma_{34b}}^2}-\frac{1}{\psi_{\Gamma_{43}}^2}\right.\nonumber\\
  & & \left. 
-\frac{\phi_{\Gamma_4}\psi_{\Gamma_3}-\phi_{\Gamma_3}\psi_{\Gamma_4}}{\psi_{\Gamma_4}^2\psi_{\Gamma_3}^2\left[\phi_{\Gamma_3}\psi_{\Gamma_4}+\phi_{\Gamma_4}\psi_{\Gamma_3} \right]}\right)\Omega_\Gamma.
\end{eqnarray}
Angle dependence is relegated to anti-cocommutativity. This is true in general \cite{Madrid}.

So let us summarize: assume we compute $\Gamma_{43}$ say.

We decompose:
$$\Gamma_{43}=\frac{1}{2}\left(p_{34}+c_{34}+\Gamma_3\Gamma_4\right).$$
The product term gives the contribution $60\zeta(3)\zeta(5)L^2$, which has its home in the symmetric square of $\mathcal{L}$, considered as an element of $\mathcal{U}(\mathcal{L})$.

$p_{34}$ gives a new period which we are waiting for. 
And $c_{34}\equiv c_{34}(\theta)\in \mathcal{L}_2$, $\notin\mathcal{L}_3$, carries all the angle dependence of $c^1_{\Gamma_{34}}$. 
This opens a vast arena of questions for algebraic geometry to be answered in the future.
It also shows how beautiful and well-organized the arena of special functions is which describes any finite order in quantum field theory.
\end{itemize}

\end{document}